\documentclass[twocolumn,english]{iopart}
\usepackage[T1]{fontenc}
\usepackage[latin9]{inputenc}
\usepackage{amstext}
\usepackage{amssymb}
\usepackage{graphicx}

\makeatletter
\usepackage{iopams}
\usepackage{setstack}

\makeatother

\usepackage{babel}
\begin{document}
\global\long\def\ket#1{\left|#1\right\rangle }

\global\long\def\bra#1{\left\langle #1\right|}

\global\long\def\braket#1#2{\left\langle #1\left|\vphantom{#1}#2\right.\right\rangle }

\global\long\def\ketbra#1#2{\left|#1\vphantom{#2}\right\rangle \left\langle \vphantom{#1}#2\right|}

\global\long\def\braOket#1#2#3{\left\langle #1\left|\vphantom{#1#3}#2\right|#3\right\rangle }

\global\long\def\mc#1{\mathcal{#1}}

\global\long\def\nrm#1{\left\Vert #1\right\Vert }

\title{The role of singular values in entanglement distillation and unambiguous
state discrimination}

\title[Singular values in entanglement distillation and unambiguous state
discrimination]{}

\author{Raam Uzdin}

\address{Racah Institute of Physics, The Hebrew University, Jerusalem, Israel.}

\ead{raam@mail.huji.ac.il}
\begin{abstract}
Unambiguous (non-orthogonal) state discrimination (USD) has a fundamental
importance in quantum information and quantum cryptography. Various
aspects of two-state and multiple-state USD are studied here using
singular value decomposition (SVD) of the evolution operator that
describes a given state discriminating system. In particular, we relate
the minimal angle between states to the ratio of the minimal and maximal
singular values. This is supported by a simple geometrical picture
in two-state USD. Furthermore, by studying the singular vectors population
we find that the minimal angle between input vectors in multiple-state
USD is always larger than the minimal angle in two-state USD in the
same system. We apply our results to study what states can undergo
entanglement distillation in a given system. 
\end{abstract}
\maketitle

\section{Introduction}

Unambiguous state discrimination (USD \cite{Barnett Book,Jaeger Book,Bergou SD rev,Chefles})
refers to a process that is capable of detecting whether a quantum
system was prepared in state $\ket{g_{i}}$, or in state \textbf{$\ket{g_{j}}$,
}where the states are not orthogonal to each other $\braket{g_{i}}{g_{j}}\neq0$,
but they are priory known. This process has prime importance in quantum
information and quantum cryptography \cite{Barnett Book,Jaeger Book,Bennett}.
In particular in this work we focus on the application of USD to entanglement
distillation (ED) \cite{Chefles}. We show that the singular values
of the ``lossy evolution operator'' that describes a given USD system,
encapsulate the USD properties of the system. We use the singular
values analysis to derive general results on USD and entanglement
distillation. 

It is well known that standard projective measurements involve an
intrinsic discrimination error that depends on the overlap of the
states \cite{Peres,Nielsen}. Yet, this error can be avoided by using
a different type of measurements called POVM (positive operator valued
measure) \cite{Peres,Nielsen}. POVM's are considered to be a part
of standard quantum theory since the Neumark dilation theorem \cite{Peres}
insures that any POVM can be implemented as a standard projective
measurement in a larger Hilbert space. While POVM's allow to have
zero discrimination error they involve an intrinsic non-zero probability
of obtaining an ``inconclusive result'' from which the input state
cannot be inferred. 

Starting from the pioneering work of Ivanovich \cite{Ivanovich},
Dieks \cite{Dieks}, and Peres \cite{Peres firsts}, USD has been
extensively studied over the years (\cite{Barnett Book,Jaeger Book,Bergou SD rev,Chefles,BB discr}
and references therein). Typically the question of interest in USD
is how to minimize the inconclusive result probability. In this work,
however, we are not interested in designing an optimal system for
a given set of input states, but to study the USD properties of a
given system. For example, in the context of ED this translates to
identifying the different states that can be distilled into maximally
entangled states by a given system. Another motivation for studying
the singular values associated with a USD task, comes from a recent
work where it is shown that the singular values of the aforementioned
lossy evolution operator determines the minimal time-energy cost needed
for implementing the USD by means of a unitary embedding.

The relation between POVM and USD to lossy evolution operators was
first presented and experimentally demonstrated in \cite{Hutnner}.
This relation has also been used in \cite{Bergou embed 2000,Bergou embed 2006,Uwe neum brach,opt and sold disc}.
Recently, a complete equivalence and its details were studied in \cite{equive RU}.
For this work it suffices to know that given a lossy evolution operator
$K$, one can find an equivalent POVM measurement \cite{equive RU}.
Conversely, given USD POVM measurement operators, it is possible to
find an equivalent $K$. We wish to emphasize that any system whose
evolution can be described by a lossy evolution operator (system or
subsystem where the norm of the states is not conserved) is capable
of USD and entanglement distillation.

After some background and preliminaries in section II, in section
III we show the key role of the singular values of $K$ in the analysis
of the USD capabilities of $K$. In particular, we study the minimal
angle between states in two-state USD. We relate it to the minimal
and maximal singular values of $K$ and provide a geometrical picture
of the process. Next we study multiple-state USD and find that the
minimal angle between states in multiple state USD, must be larger
than the minimal angle in two-state USD. While in section III we investigated
the relation between a given input set of states and the discriminating
operator K, in section IV we discuss the relation between different
set of states that K discriminates unambiguously. This has practical
importance to USD based entanglement distillation, as it relates all
the state that can be distilled by a given system.

\section{Background and preliminaries}

\subsection{Lossy evolution operator and POVM}

Let $\ket{\psi(t)}\in\mathbb{C}^{N}$ be a state in an $N$-level
system. The evolution of the state from $t=t_{i}$ to $t=t_{_{f}}$
generated by a lossy evolution operator is given by:
\begin{eqnarray}
\ket{\psi(t_{f})} & =K & \ket{\psi(t_{i})},\\
K^{\dagger}K & \neq & I.\label{eq: K NU}
\end{eqnarray}
where $K\in\mathbb{C}^{N\times N}$. Accordingly, the density matrix
of the state evolves according to $K\ketbra{\psi}{\psi}K^{\dagger}$.
The main difference in comparison to Kraus map evolution $\sum_{i}M_{i}\ketbra{\psi}{\psi}M_{i}^{\dagger}$
is that it does not conserve the trace of the density matrix (see
(\ref{eq: K NU})). Note that if $\ket{\psi}$ describes only a subsystem
(e.g. there other levels that are not accounted for in $\ket{\psi}$
), then the trace of the density matrix of the subsystem need not
be conserved. Hence, we call this ``lossy'' evolution as in Ref.
\cite{Hutnner}. Nevertheless, the loss mechanism does not have to
be related to absorption. In quantum mechanics it can be due to tunneling
or ionization that lead to probability loss in the subsystem of interest.
There are two ways to generate such a $K$. The first is to use Schrödinger
Eq. with some non-Hermitian Hamiltonian that takes losses into account.
This formalism is very useful for the description of resonances and
metastable states \cite{Nim book}. The other way to generate $K$
is by embedding it in a larger Hilbert space (e.g. by coupling the
system to an ancilla) where a unitary evolution takes place \cite{Hutnner,Bergou embed 2000,Bergou embed 2006,Uwe neum brach,opt and sold disc}.
For characteristics properties of such embeddings see \cite{embedding cost,Popescu}.

\subsection{Singular value decomposition and the spectral norm}

Let $K\in\mathbb{C}^{N\times N}$ be a general linear operator in
a Hilbert space of dimension $N$. According to the singular value
decomposition (SVD \cite{Horn}) $K$ can always be written as:
\begin{equation}
K=\sum s_{i}\ketbra{u_{i}}{v_{i}},\label{eq: bracket svd}
\end{equation}
where $s_{i}\ge0$ are called the singular values of $K$, and the
vectors $\ket{v_{i}},\ket{u_{i}}$ satisfy 
\begin{eqnarray}
K\ket{v_{i}} & = & s_{i}\ket{u_{i}},\label{eq: svd k*v=00003Dsu}\\
K^{\dagger}\ket{u_{i}} & = & s_{i}\ket{v_{i}}.\label{eq: svd kd*u=00003Dsv}
\end{eqnarray}
While $\braket{v_{i}}{v_{j}}=\delta_{ij}$ and $\braket{u_{i}}{u_{j}}=\delta_{ij}$,
the overlap $\braket{u_{i}}{v_{j\neq i}}$ is in general not zero.
The singular values and singular vector are calculated using:

\begin{eqnarray}
K^{\dagger}K\ket{v_{i}} & = & s_{i}^{2}\ket{v_{i}},\label{eq: Kd K svd}\\
KK^{\dagger}\ket{u_{i}} & = & s_{i}^{2}\ket{u_{i}}.
\end{eqnarray}
Finally, the singular values induce a few important matrix norms.
In this work, however, we shall always use the spectral norm \cite{Horn}:
\begin{equation}
\nrm K=\max(s_{i})=\sqrt{\max(\text{eigenvalues}(K^{\dagger}K))}.
\end{equation}
Finally we comment that in systems with no gain (in contrast to amplifying
medium in optics) $K$ can never increase the amplitude of a state.
The passiveness of $K$ is given by the condition \cite{NU resources}:
\begin{equation}
\nrm{K_{\text{passive}}}\le1.\label{eq: passiveness cond}
\end{equation}

\subsection{Structure of a lossy USD evolution operator}

In this subsection we write the general structure of a transformation
that takes a non-orthogonal input set of states to a set of orthogonal
output states. Let $G$ be a column matrix of the linearly-independent
\cite{lin indep Chefles}, non-orthogonal, non-normalized input vectors
$\ket{g_{i}}$ we wish to discriminate, and let $G_{\perp}$ be the
columns matrix of the of the bi-orthogonal vectors $\ket{g_{i}^{\perp}}$
so that:
\begin{equation}
G_{\perp}^{\dagger}G=I.
\end{equation}
Clearly, $G_{\perp}^{\dagger}=G^{-1}$. $G$ is invertible due to
the linear independence of its columns. A discriminating $K$ has
the general form:
\begin{eqnarray}
K & = & U_{out}\Lambda G^{-1},\label{eq: K matrix}
\end{eqnarray}
where $\Lambda$ is a diagonal matrix that affects the posterior probability
to detect the vectors, and $U_{\text{out}}$$ $ is a unitary matrix
that determines the basis in which the results are expressed (the
measurement basis). Without it, the results will appear in the computational
basis (1,0,0..), (0,1,0,...) and so on. The choice $\Lambda_{ii}=\text{const}$
is of particular importance as it used for ED as we show in the next
subsection. Equation (\ref{eq: K matrix}) can also be written as:
\begin{equation}
K=\sum_{i=1}^{N}\Lambda_{ii}\ketbra{\psi_{i}}{g_{i}^{\perp}},\label{eq: K USD braket}
\end{equation}
where $\braket{g_{i}^{\perp}}{g_{j}}=\delta_{ij}$ and $\ket{\psi_{i}}$
are the columns of $U_{out}$. This transformation implements $\ket{g_{1}}\overset{K}{\to}\Lambda_{ii}\ket{\psi_{i}}$.
Note that $\ket{g_{i}^{\perp}}$ are not orthogonal to each other
so (\ref{eq: K USD braket}) should not be confused with singular
value decomposition (\ref{eq: bracket svd}).

\subsection{USD based entanglement distillation\label{sub: USD-based-entanglement}}

Entanglement distillation (ED) refers to a process where a pure entangled
bipartite state is converted into a maximally entangled state by applying
only local operations. According to \cite{Majorization Nielsen} this
cannot be done deterministically. There is some probability of success
that depends on the needed entanglement increase. As described in
detail by Chefles \cite{Chefles} ED can be implemented by USD. Although
to a large extent this subsection repeats the analysis of Chefles
we find it worthwhile to repeat it in our notation and to use the
lossy evolution operator point of view.

Consider an entangled state constructed from non-orthogonal states
$\braket{x_{j}}{x_{k\neq j}}\neq0,\braket{y_{j}}{y_{k\neq j}}\neq0$:
\begin{equation}
\ket{\Psi}=\sum_{i=1}^{N}c_{k}\ket{x_{k}}_{A}\otimes\ket{y_{k}}_{B}.
\end{equation}
Our goal is to find a local transformation on side A that will turn
this state into a maximally entangled state of the form: $\ket{\psi_{max}}=b\sum_{k=1}^{N}\ket{\phi_{k}}\otimes\ket{\varphi_{k}}$
where $\{\ket{\phi_{k}}\}_{k=1}^{N}$ and $\{\ket{\varphi_{k}}\}_{k=1}^{N}$
are some orthonormal bases and $b<1/\sqrt{N}$. We start by spanning
the $\ket{y_{k}}$ in terms of some other orthonormal basis $\{\ket{\varphi_{k}}\}_{k=1}^{N}$
:
\begin{equation}
\ket{\Psi}=\sum_{k=1}^{N}c_{k}\ket{x_{k}}_{A}\otimes\sum_{i}d_{ki}\ket{\varphi_{i}}_{B}=\sum_{i=1}^{N}\ket{g_{k}}_{A}\otimes\ket{\varphi_{i}}_{B},
\end{equation}
where the $\ket{g_{i}}=\sum_{k=1}^{N}c_{k}d_{ki}\ket{x_{k}}$ states
are non-orthogonal and non-normalized. Now we operate locally on system
$A$ with the lossy evolution operator:
\begin{equation}
K_{A}=\frac{G^{-1}}{\nrm{G^{-1}}},
\end{equation}
where, as before, $G$ is a matrix whose columns are given by the
$\ket{g_{i}}$ states. The normalization factor $1/\nrm{G^{-1}}$
insures that $K_{A}$ is a passive operator (see (\ref{eq: passiveness cond})).
Operating with $K_{A}$ on $\ket{\Psi}$ we get:
\begin{equation}
K_{A}\ket{\Psi}=\frac{1}{\nrm{G^{-1}}}\sum_{i=1}^{N}\ket{e_{k}}_{A}\otimes\ket{\varphi_{i}}_{B},
\end{equation}
where $\ket{e_{k}}$ is the computational basis. The distillation
success probability is given by the square of the state's norm: 
\begin{equation}
P_{dist}=\frac{N}{\nrm{G^{-1}}^{2}}.
\end{equation}
Like in USD, for ED we also use a lossy evolution operator that turns
a non-orthogonal set of states into an orthogonal set. The difference
is that in the USD setup the operator works on a single input state
in each experiment, while in ED $K$ operates on all the input states
simultaneously in each experiment. Another difference is that for
ED we insist that the output vectors will have the same weights (i.e.
that $\Lambda_{ii}=\text{const}$ in (\ref{eq: K USD braket})). 

Before moving on, another relation between USD and ED should be mentioned.
A system which is prepared in one of the \textit{non-orthogonal} normalized
states $\{\ket{h_{i}}\}_{i=1}^{N}$ with probability $p_{i}$ is described
by the density matrix: $\rho_{\text{USD}}=\sum_{i=1}^{N}p_{i}\ketbra{h_{i}}{h_{i}}$.
This positive matrix has an eigenvalue decomposition: 
\begin{equation}
\rho_{\text{USD}}=\sum_{i=1}^{N}\sigma_{i}\ketbra{\varphi_{i}}{\varphi_{i}},\label{eq: rho USD}
\end{equation}
where $0\le\sigma_{i}\le1$. Next we point out that any pure bipartite
state can be written in the Schmidt from \cite{Nielsen}:
\begin{equation}
\ket{\Psi}=\sum_{i=1}^{N}\lambda_{i}\ket{\xi_{i}}_{A}\otimes\ket{\chi_{i}}_{B},
\end{equation}
where $\{\ket{\xi_{i}}_{A}\}_{i=1}^{N}$ and $\{\ket{\chi_{i}}_{A}\}_{i=1}^{N}$
are orthogonal bases and the coefficients $0\le\lambda_{i}\le1$ are
the Schmidt coefficients. The reduced density matrix of system A (B)
is obtained by taking the partial trace on $B$ ($A$) so that:
\begin{equation}
\rho_{A}=\sum_{i=1}^{N}\lambda_{i}\ket{\xi_{i}}_{A}\bra{\xi_{i}}_{A},
\end{equation}
which has the same form as (\ref{eq: rho USD}). Therefore the eigenvalues
of the density matrix in USD play the same role as the Schmidt coefficients
in pure entangled states. This analogue will become useful in section
IV. $ $

Although we selected ED as our leading physical example our findings
are relevant to any USD application.

\section{Singular values and unambiguous state discrimination}

Before explicitly studying the role of singular values in USD we wish
to present a general argument why singular values capture the essence
of USD. We shall do so by comparing two systems $A$ and $B$ (not
to be confused with $A$ and $B$ of section \ref{sub: USD-based-entanglement})
and checking when they are ``USD equivalent''.

\subsection{USD equivalent systems}

Consider a lossy evolution operator $K_{A}$ that transform anon-orthogonal
set of states in system $A$ $\{\ket{g_{i}}_{A}\}_{i=1}^{N}$ to orthogonal
states $\{\ket{\psi_{i}}_{A}\}_{i=1}^{N}$ (\ref{eq: K USD braket}).
Let the lossy evolution operator in system $B$, be $K_{B}=K_{A}U_{R}$
where $U_{R}$ is unitary matrix. Clearly the operator $K_{B}$ will
discriminate the set $\{\ket{g_{i}}_{B}\}_{i=1}^{N}=\{U_{R}^{\dagger}\ket{g_{i}}_{A}\}_{i=1}^{N}$.
This set has exactly the same properties as $\{\ket{g_{i}}_{A}\}_{i=1}^{N}$.
The angles and inner products between all the states are exactly the
same. Hence $K_{A}$ and $K_{B}$ have the same discrimination properties.
Multiplying of $K_{B}$ from the left by $U_{L}$ simply change the
output measurement basis to $\{\ket{\psi_{i}}_{B}\}_{i=1}^{N}=\{U_{L}\ket{\psi_{i}}_{A}\}_{i=1}^{N}$
but it does not affect the discrimination properties. In conclusion,
since $K_{A}$ and $K_{B}=U_{L}K_{A}U_{R}$ have the exactly the same
discrimination properties we call them USD equivalent. What quantities
are invariant under such double unitary transformations? From (\ref{eq: bracket svd})
or (\ref{eq: Kd K svd}) it is easy to see that the singular values
are invariant under any such transformation. Consequently, two systems
are USD equivalent if they have the same singular values. Furthermore,
we expect that any discrimination property that depends only on $K$
and not on the input states will be a function of the singular values.
For example, in the next section we study the minimal discrimination
angle between input states and express it in term of the singular
values. It should be noted, though, that not all quantities of interested
in USD are input independent. The success or detection probability
does depend on the input%
\footnote{For example it depends on the prior probability of each input state.
This information is contained in the input density matrix not in K.%
} so one can not expect an exact expression for it using only the singular
values of $K$. 

In the context of ED, USD equivalence means that the set input states
that can be distilled by $A$, is related to the distillable input
states of $B$ by a local unitary transformation. Furthermore, the
maximally entangled output states of $A$ are related to those of
$B$ by another local unitary transformation.

After discussing the equivalence of two systems we now turn to study
the set of states that can be discriminated by a given system $K$.

\[
\]

\subsection{The basic relation between singular values and USD}

In this section we show that in the most basic USD example in Hilbert
space of dimension two (two levels or one qubit), the angle between
the non-orthogonal states that can be discriminated by $K$ is a function
of the singular values ratio of $K$. In the next section we generalize
this and show that this example is also important in multiple-state
USD. For two-level systems the SVD of the lossy evolution operator
(\ref{eq: bracket svd}) is: 
\begin{eqnarray}
K=K_{2\times2} & = & s_{min}\ketbra{u_{min}}{v_{min}}+s_{max}\ketbra{u_{max}}{v_{max}}.
\end{eqnarray}
Consider now the special input vectors: 
\begin{equation}
\ket{g_{\pm}}=s_{min}\ket{v_{max}}\pm s_{max}\ket{v_{min}}.\label{eq: g+/-}
\end{equation}
The corresponding output vectors are:
\begin{equation}
K\ket{g_{\pm}}=s_{min}s_{max}(\ket{u_{max}}\pm\ket{u_{min}}).\label{eq: K g+/-}
\end{equation}
Since the $\ket{u_{i}}'s$ are orthonormal, we get that the output
vectors are orthogonal while the input vectors are not. The input
angle is:

\begin{equation}
\cos\theta_{0}=\frac{\left|\braket{g_{-}}{g_{+}}\right|}{\sqrt{\left|\braket{g_{+}}{g_{+}}\right|\left|\braket{g_{-}}{g_{-}}\right|}}=\frac{1-\left(\frac{s_{min}}{s_{max}}\right)^{2}}{1+\left(\frac{s_{min}}{s_{max}}\right)^{2}}.\label{eq: simple cos}
\end{equation}
Figure 1 offers a geometrical interpretation that differs from the
standard geometrical interpretation of USD. Usually USD is explained
by a unitary rotation of the vectors in a higher dimensional Hilbert
space (e.g. \cite{opt real myers brandt 97}). Here, the vectors remain
in the same space but different axes are squeezed by different factors.
In figure 1 the non-orthogonal thin-blue vectors undergo a nonuniform
stretching. The x axis is squeezed by a factor $s_{min}<1$, and the
y axis is stretched by a factor $s_{max}>1$ (for the illustration
we used $s_{max}>1$ but for passive systems $s_{max}\le1$). Due
to the anisotropic stretching, the resulting thick-red vectors are
orthogonal to each other.

An easier way to obtain (\ref{eq: simple cos}) follows immediately
from Fig. 1. According to the figure: $\cos\frac{\theta_{0}}{2}=\frac{s_{max}}{\sqrt{s_{min}^{2}+s_{max}^{2}}}$.
Squaring it and using half angle formula leads directly to (\ref{eq: simple cos}).
The fact that the vectors may be complex does not matter, as will
be explained in the next section. Alternatively, (\ref{eq: simple cos})
can be written in a simpler way using half the angle:

\begin{equation}
\tan\frac{\theta_{0}}{2}=\frac{s_{min}}{s_{max}}=\frac{1}{\nrm K\nrm{K^{-1}}}.\label{eq: simple tan}
\end{equation}

\begin{figure}
\begin{centering}
\includegraphics[width=12cm]{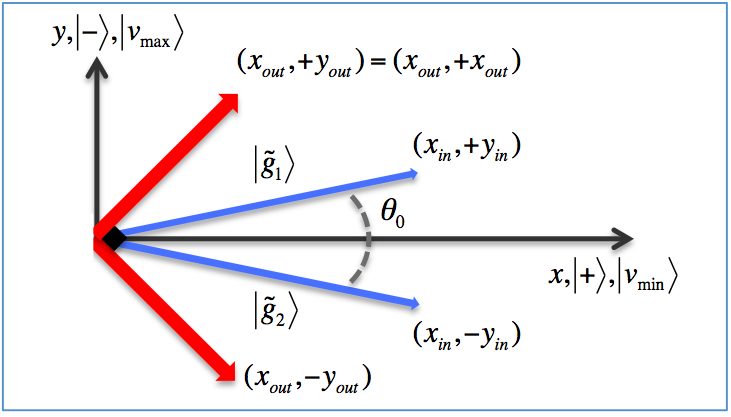}
\par\end{centering}

\caption{(color online) Singular values geometrical interpretation of two states
USD. By a unitary rotation the two-state discrimination problem in
an arbitrary size Hilbert space can be reduced to 2D real vectors
transformation as depicted above. The $x$ and $y$ component of the
non-orthogonal blue vectors are squeezed by different factors so that
after the squeezing the two vectors become orthogonal. The smallest
discrimination angle is determined by the smallest and largest singular
values of the transformation.}
\end{figure}
We comment that using the singular values approach and expressions
(\ref{eq: g+/-}),(\ref{eq: K g+/-}) it is straightforward to obtain
the known result of the detection probability for optimal two-state
discrimination (\cite{Dieks} or p. 154 of \cite{Jaeger Book}): $P_{D}=1-\left|\braket{g_{i}}{g_{j}}\right|$.
In the next section we study two-state discrimination in a scenario
of multiple USD.

\subsection{The best discrimination between two states}

In a larger Hilbert space, are there two input vectors $\ket{g_{1,2}}$
that lead to a smaller discrimination angle than $\theta_{0}$ obtained
in the previous section? To answer this we define the sum-difference
basis:
\begin{eqnarray}
\ket{\pm} & = & \frac{\ket{g_{1}}\pm\ket{g_{2}}}{\sqrt{1+2\braket{g_{2}}{g_{1}}}},\label{eq: pm basis}\\
\braket{g_{2}}{g_{1}} & > & 0.\label{eq: a1a2 cond}
\end{eqnarray}
Without loss of generality we assumed the states are normalized ($\braket{g_{1}}{g_{1}}=\braket{g_{2}}{g_{2}}=1)$
to simplify the writing. Condition (\ref{eq: a1a2 cond}) ensures
the orthogonality of this basis $\braket -+=0$. One can always change
the relative global phase of the states so that (\ref{eq: a1a2 cond})
holds%
\footnote{this condition can be relaxed, but then several phase factors must
be included in the definition of the $\ket{\pm}$ basis (\ref{eq: pm basis}).%
}. In this new basis, the original input complex $N$-component vectors
$\ket{g_{1,2}}$ have the form:

\begin{equation}
\ket{\tilde{g}_{1,2}}=(x_{\text{in}},\pm y_{\text{in}}),
\end{equation}
where $x_{\text{in}},y_{\text{in}}$ are \textit{real} numbers and
the tilde denote representation in the $\ket{\pm}$ basis. These are
the coefficients in the new basis, but one can apply a unitary transformation
$U_{R}$ and rotate $\ket{\pm}$ to the computational basis $(1,0)$
and $(0,1)$. By applying the same type of rotation $U_{L}$ to the
output vectors we have:

\begin{equation}
\tilde{K}\ket{\tilde{g}_{1,2}}=(x_{\text{out}},\pm y_{\text{out}}),
\end{equation}
where $\tilde{K}=U_{L}KU_{R}^{\dagger}$ . As explained before, such
unitary rotations have no impact on the discrimination properties
of $K$. That is, $\tilde{K}$ is USD equivalent to $K$ . If so,
the problem of two-state discrimination in $N$ dimensional Hilbert
space was reduced to two-dimensional \textit{real} vector space. In
USD we want the output vectors to be orthogonal so:

\begin{equation}
\tilde{K}\ket{\tilde{g}_{1,2}}=(x_{\text{out}},\pm x_{\text{out}}).
\end{equation}
The input angle satisfies:
\begin{equation}
\tan\frac{\phi}{2}=\frac{y_{\text{in}}/x_{\text{in}}}{x_{\text{out}}/x_{\text{out}}}=\frac{y_{\text{in}}/x_{\text{in}}}{y_{\text{out}}/x_{\text{out}}}=\frac{x_{\text{out}}}{x_{\text{in}}}\frac{y_{\text{in}}}{y_{\text{out}}}.
\end{equation}
Since $\frac{x_{\text{out}}}{x_{\text{in}}}\ge s_{\text{min}}$ and
$\frac{y_{\text{in}}}{y_{\text{out}}}\ge1/s_{\text{max}}$ we get:

\begin{equation}
\tan\frac{\phi}{2}\ge\frac{s_{\text{min}}}{s_{\text{max}}}.
\end{equation}
Comparing to (\ref{eq: simple tan}) we see that an equality takes
place when $\phi=\theta_{0}.$ Therefore, the best discrimination
angle is the one obtained in the previous section. The input states
that achieve this angle are given by (\ref{eq: g+/-}). We conclude
that the best discrimination of $K$ can generate is given by:
\begin{equation}
\theta_{\text{best}}=2\arctan\frac{s_{\text{min}}}{s_{\text{max}}}=2\arctan\frac{1}{\nrm K_{sp}\nrm{K^{-1}}_{sp}}.\label{eq: best arctan}
\end{equation}
After some algebra and using $b=1$ and $a=2$ in formula (1.7) of
\cite{arcsin ineq} we obtain:

\begin{equation}
\frac{3}{2}\frac{1}{\nrm K_{sp}\nrm{K^{-1}}_{sp}}\le\theta_{\text{best}}\le\frac{2}{\nrm K_{sp}\nrm{K^{-1}}_{sp}}.
\end{equation}
Significantly tighter inequalities can be written by making other
choices of $a$ and $b$, but here we prefer to present the simplest
expressions. 

Interestingly, this result carries over directly to the inverse problem
of ``non-Hermitian cooling'' \cite{NU resources}. There, the goal
is to cool a statistical mixture of orthogonal states by making them
more parallel to each other. The best discrimination angle of $K$
is the best cooling angle of $K^{-1}$. However formula (\ref{eq: best arctan})
is invariant to $K\leftrightarrow K^{-1}$ transformation, so the
best discrimination angle of $K$ is also the best cooling angle of
$K$. 

In the next section we show why typically in USD of more than two
non-orthogonal states this angle cannot be achieved.

\subsection{Minimal angle in multiple-state USD and singular vector population}

Let $\{\ket{\alpha_{i}}\}_{i=1}^{N}$ be a ``completely non-orthogonal''
set of states ( $\braket{\alpha_{i}}{\alpha_{j}}\neq0$ for \textit{all}
$i,j$) and let $K$ be the lossy evolution operator that discriminates
them. Furthermore, let us make an ansatz that a subset $\{\ket{\alpha_{k}}\}_{k=1}^{L}$
is spanned by $L<N$ singular vectors:
\begin{equation}
\ket{\alpha_{k}}=\sum_{i=1}^{L}a_{ki}\ket{v_{i}},\label{eq: v ansatz}
\end{equation}
where the coefficients $a_{ki}$ form an invertible matrix. An important
observation on the singular values population will follow by showing
that this last ansatz contradicts the complete non-orthogonality ansatz.
Since the output states $\ket{\beta_{i}}=K\ket{\alpha_{i}}=\sum_{i=1}^{L}s_{i}a_{ki}\ket{u_{i}}$
span the image subspace $\{u_{i}\}_{i=1}^{L}$ (the $\ket{\beta_{i}}$
are linearly independent) and $\braket{\beta_{k'>L}}{\beta_{k\le L}}=0$
we get that:
\begin{equation}
0=\braket{\beta_{k'>L}}{u_{k\le L}}=s_{k}\braket{\alpha_{k'>L}}{v_{k\le L}},
\end{equation}
where we used (\ref{eq: svd kd*u=00003Dsv}). Using this in our second
ansatz (\ref{eq: v ansatz}) one obtains:
\begin{equation}
\braket{\alpha_{k'>L}}{\alpha_{k\le L}}=0.
\end{equation}
That is, if some $L$ input states are spanned by $L$ singular vectors,
all other vectors that can be discriminated must be initially orthogonal
to the aforementioned $L$ vectors. This contradicts the first complete
non-orthogonality ansatz. The implication is that if all the input
states are non-orthogonal to each other, each input vector must populate
$all$ the singular vectors $\ket{v_{i}}$: $\braket{\alpha_{k}}{v_{i}}\neq0\:\forall\: k,i$.
Another implication is that in $N>2$ USD for which $\braket{\alpha_{i}}{\alpha_{j\neq i}}\neq0$,
the smallest angle between vectors satisfies:$ $
\begin{equation}
\theta_{\text{min}}^{\text{mult}}>\theta_{\text{best}}=2\arctan\frac{s_{\text{min}}}{s_{\text{max}}}.
\end{equation}
The strong inequality follows from the fact that the minimal and maximal
singular vectors can not be exclusively populated in a completely
non-orthogaonal multiple-state USD setup. Alternatively stated, in
multiple-state USD the minimal discrimination angle can never be as
small as the optimal two-state discrimination in the same system.
The optimal two-state discrimination states are given by (\ref{eq: g+/-}).

\section{Families of discriminable states and singular value degeneracy}

In this section we ask what states can be discriminated by a given
lossy evolution $K$. In ED this translates to the question which
states can be distilled by a given USD system. Our approach is based
on finding transformations that are applied to the input states and
leave the output states orthogonal to each other. These transformations
do not have to be carried out in practice, so there is no reason they
should be unitary. In fact, we shall explore both unitary and non-unitary
transformations. The unitary transformations will be used to find
sets of discriminable states that have the same USD density matrix
eigenvalues (or the same Schmidt coefficients in ED). The non-unitary
transformation will be useful to move from one family (described by
density matrix eigenvalues/Schmidt coefficients) to a different family.

\subsection{Special unitary transformations}

Let us write the states of an input set $\{\ket{g_{k}}\}_{k=1}^{N}$
in term of the singular vectors of $K$ (\ref{eq: bracket svd}):

\begin{equation}
\ket{g_{k}}=\sum a_{ki}\ket{v_{i}}.
\end{equation}
The overlap of the output states $\ket{h_{j}}=K\ket{g_{j}}$ is:

\begin{equation}
\braket{h_{k}}{h_{j}}=\braOket{g_{k}}{K^{\dagger}K}{g_{i}}=\sum s_{i}^{2}a_{ki}^{*}a_{ji}.
\end{equation}
This overlap will not change if the following transformation is applied
to the input states:
\begin{equation}
\ket{\tilde{g}_{k}}=W\ket{g_{k}}=(\sum_{i}e^{i\phi_{i}}\ketbra{v_{i}}{v_{i}})\ket{g_{k}}.
\end{equation}
Although the output vectors are modified, they remain orthogonal to
each other. Notice that $W$ is a unitary matrix, but a very specific
one. Its eigenvectors are the $\ket{v_{i}}$ singular vectors of $K$.
If the set $\{\ket{g_{i}}\}$ can be discriminated, the set $\{W\ket{g_{i}}\}$
can be discriminated as well by the same $K$. This should not be
confused with the earlier discussion about USD equivalence of two
systems. Here $K$ is set, and we ask what are the families of states
that can be discriminated by this K. 

The orthogonal output vectors will appear in a different basis than
$\ket{h_{i}}$. Using (\ref{eq: K USD braket}) one can easily see
that the new output basis is related to the old one by:

\begin{equation}
\ket{\tilde{h}_{k}}=(\sum_{i}e^{i\phi_{i}}\ketbra{u_{i}}{u_{i}})\ket{h_{k}}.
\end{equation}
Strictly speaking when the output basis is change the POVM operators
are modified as well \cite{equive RU}. However, in embedding realizations
(\cite{embedding cost} and also \cite{Bergou embed 2000,Bergou embed 2006,opt real myers brandt 97,opt and sold disc}),
the change of basis corresponds to a unitary rotation on the ``system''
levels (i.e. without the ancilla levels)$ $ after the embedding part
has been completed. So physically the $\{\ket{\tilde{g}_{k}}\}$ states
describe family of discriminable states related to the same device
even though the POVM operator are modified by $W$. 

Positive $K$ operators have a special property. In \cite{embedding cost}
it was shown that the positivity is a necessary condition for minimizing
the time-energy resources needed for unitary embedding of the desired
USD. Note that if $K$ is not positive, it can be made positive by
applying a unitary from the left. Since positive $K$ and $W$ have
the same \textit{eigenvectors} they commute. This means that the $W$
rotation can be performed before or after the distillation/USD and
the result will be the same.

\subsection{Singular value degeneracy}

It may happen that two or more singular vectors have the same singular
values. Such degeneracies appear automatically in unitary embedding
realizations of lossy evolution when the ancilla dimension is smaller
than the system dimension by more than one level \cite{embedding cost}.
In the presence of singular value degeneracy, on top of $W$, it is
possible to apply any unitary $V$ that mixes the degenerate singular
vectors. 

When the degenerate singular values are equal to one (as in the unitary
embedding scheme \cite{embedding cost}) there is an interesting physical
consequence. In the case of an inconclusive result, the degeneracy
determines the rank of the density matrix after the measurement. The
inconclusive density matrix is $\rho_{?}=M_{?}\rho M_{?}^{\dagger}$
where $M_{?}=\sqrt{I-K^{\dagger}K}$ (see \cite{equive RU}). In the
extreme case where $N-1$ singular values are equal to one, the rank
of $ $$\rho_{?}$ is one. This means that the remaining density matrix
contains no information at all. If an inconclusive result is obtained,
the state of the system is always the same regardless of the input.
This case naturally appears in the embedding scheme when the ancilla
has only one level \cite{embedding cost}. This is consistent with
the fact that no information can be encoded in a single quantum level.

The local unitary transformation $W$ and $V$ do not change the entanglement
of the original state (Schmidt coefficients are invariant to local
unitary transformations). Alternatively in USD, $W$ and $V$ do not
change the eigenvalues of the density matrix. In what follows we describe
a non-unitary transformation that change the Schmidt coefficients
and extend the family of states that can be distilled/discriminated
by a given system.

\subsection{The entanglement distillation transformation}

Given $K$, it is clear that the columns of $G=K^{-1}$ describe a
possible set of vectors that $K$ can discriminate since $I=KK^{-1}$.
This corresponds to the case $\Lambda_{ii}=\text{const}$ (see (\ref{eq: K USD braket}))
that is used for distillation. Multiplying both sides from the right
by a unitary matrix $U_{0}$ we get:
\begin{equation}
U_{0}=KK^{-1}U_{0}=K\tilde{G},
\end{equation}
where:

\begin{equation}
\tilde{G}=K^{-1}U_{0}.
\end{equation}
$\tilde{G}$ describes a new set of input states (column vectors)
that can be discriminated by $K$. Notice that unitary transformation
acting from the right is not a unitary rotation for the columns of
$G$. Therefore, in contrast to the previous unitary transformations,
here the relative angles between the new input states $\tilde{G}$
are different for different choices of $U_{0}$. Since this transformation
modifies the Schmidt coefficients it extends the family of states
that can be distilled with respect the family formed by $W$ and $V$
.

\section{Concluding remarks}

This work shows that various insights to USD can be gained from singular
value analysis of the lossy evolution operator. It is likely that
deriving the same findings directly from the POVM operators and without
singular values would prove to be rather difficult. It is interesting
to see what other USD and entanglements distillation features can
be unraveled from the singular value analysis of the lossy evolution
operator.

\ack{}{The author is in debt to Omri Gat for stimulating discussions.\protect \\
}

\end{document}